\documentclass{jfm}

\usepackage{graphicx}
\usepackage{newtxtext}
\usepackage{newtxmath}
\usepackage{natbib}
\usepackage{hyperref}
\hypersetup{
    colorlinks = true,
    urlcolor   = blue,
    citecolor  = black,
}

\newcommand{\RomanNumeralCaps}[1]

\usepackage[justification=justified]{caption}

\newcommand\Arch{\mbox{\textit{Ar}}}
\usepackage{bm}

\title{On the rising and sinking of granular bubbles and droplets}

\author{Jens P. Metzger\aff{1},
  Louis Girardin\aff{2},
  Nicholas A. Conzelmann\aff{1}
 \and Christoph R. M{\"u}ller\aff{1}
    \corresp{\email{muelchri@ethz.ch}}}

\affiliation{\aff{1}Department of Mechanical and Process Engineering, ETH Zurich, 8092 Zurich, CH
\aff{2}Department of Mechanical Engineering, University College of London, W1W 7TS London, UK}

\begin{document}
\maketitle

\begin{abstract}
Recently, the existence of so-called granular bubbles and droplets has been demonstrated experimentally. Granular bubbles and droplets are clusters of particles that respectively rise and sink if submerged in an aerated and vibrated bed of another granular material of different size and/or density. However, currently there is no model that explains the coherent motion of these clusters and predicts the transition between a rising and sinking motion. Here, we propose an analytical model predicting accurately the neutral buoyancy limit of a granular bubble/droplet. This model allows the compilation of a regime map identifying five distinct regimes of granular bubble/droplet motion.
\end{abstract}



\section{Introduction}
\label{sec:intro}
\begin{figure}
\centerline{\includegraphics[width=\linewidth]{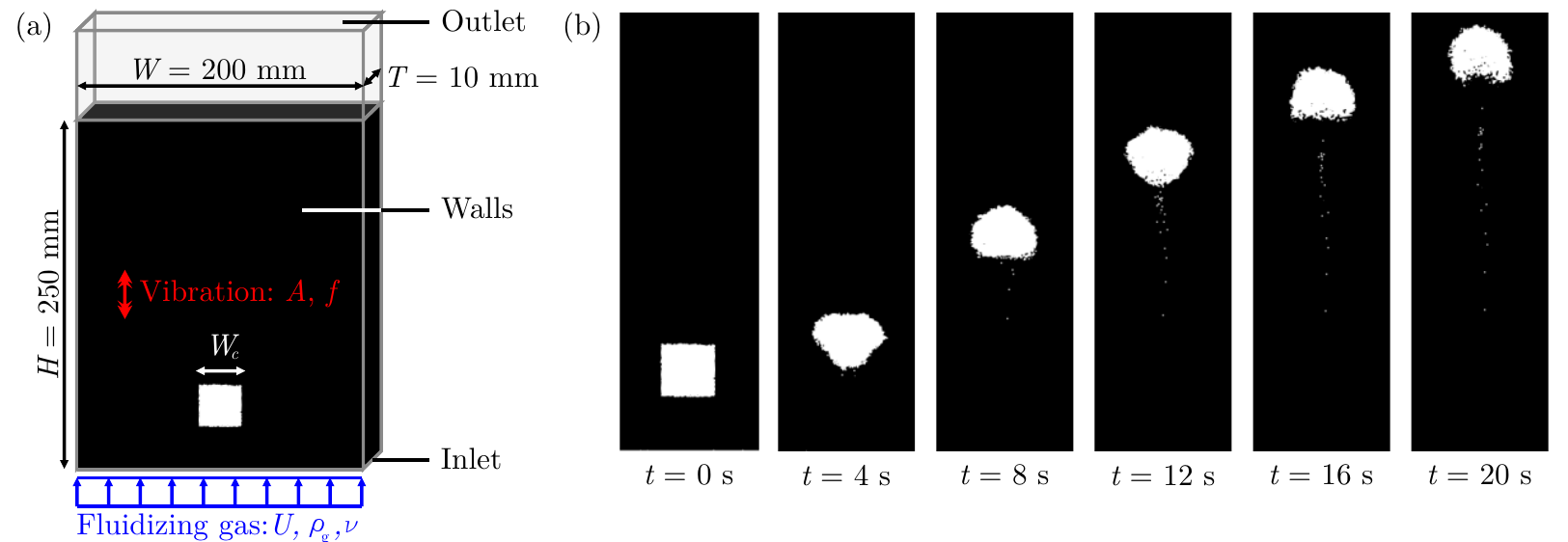}}
\captionsetup{width=\linewidth} 
\caption{\label{fig:TimeSeries}(a) Numerical setup: Bulk particles (black,  $d_b = 1.16 \text{ mm}$ and $\rho_b = 6000 \text{ kg/m}^3$) are filled to a height $H$ in a container of width $W$ and depth $T$. Granular cluster particles (white) are initialized as a square-cuboid ($W_c = 30$~mm) that is immersed in the bulk phase. The bed is subjected to a vertical, sinusoidal vibration ($A = 1 \text{ mm}$ and $f = 10 \text{ Hz}$) and an upwards gas flow $U = 1.13 \text{ m/s}$ with $\rho_g = 1.2 \text{~kg/m\textsuperscript{3}}$ and $\nu = 1.5\times10^{-5} \text{ m}^2/\text{s}$. (b) Time series of a rising granular bubble formed by granular cluster particles ($d_c = 1.45 \text{ mm}$ and $\rho_c = 3000 \text{ kg/m}^3$). The images show a central cut-out of the bed with a width of $0.375W$ and height $H$.}
\end{figure}
Granular materials are complex systems that can exhibit liquid-like behavior under agitation. Examples are gas bubbles in gas-solid fluidized beds \citep{RN1}, breaking waves in granular shear layers \citep{RN2}, gas-solid fingering patterns akin to Rayleigh-Taylor instabilities \citep{RN3}, and condensation-like droplet formation of particles on tapped plates \citep{RN4}. Very recently, \citet{RN5} have observed further liquid-like phenomena in binary granular materials that are subjected to simultaneous vibration and aeration. When a cluster composed of particles with diameter $d_c$ and density $\rho_c$ is immersed in a bed of particles of different diameter $d_b$ and density $\rho_b$, the cluster forms a coherent structure rising similar to a gas bubble in a liquid. Such clusters have been termed "granular bubbles" (see figure~\ref{fig:TimeSeries}~(b)) and have been observed to form for $d_c/d_b > 1$ and $\rho_c/\rho_b <1$ albeit the absence of surface tension at the cluster interface. Conversely, for $d_c/d_b <1 $ and $\rho_c/\rho_b > 1$ the cluster (termed “granular droplet”) sinks and splits similar to a droplet of dense liquid falling in a miscible but lighter liquid. Despite some similarities with their fluid analogues, the mechanism driving granular bubbles and droplets must differ appreciably from their liquid counterparts as granular materials readily solidify under pressure and lose their fluidity due to frictional forces. \citet{RN5} have argued that a granular bubble rises for $d_c > d_b$ as gas required to fluidize the granular material is drawn into the bubble, counteracting gravity through an increased drag. On the other hand, a granular droplet sinks for $d_c < d_b$ as gas bypasses the droplet. As the work of \citet{RN5} was largely experimental relying on only two different ratios for $d_c/d_b$ and $\rho_c/\rho_b$, there is still very little understanding of the underlying physics of these new phenomena. Here, we derive an analytical model that predicts the neutral buoyancy limit of a granular cluster allowing the construction of a regime map. This regime map reveals the existence of three additional, previously unreported, regimes.

\section{Numerical Setup}
\label{sec:Methods}

A critical assessment of the hypothesis of \citet{RN5} concerning the transition between rising and sinking granular clusters requires quantitative information on the gas flow near a cluster. However, this information is not easily accessible by experiments due to the opaque nature of granular matter. To this end computational fluid dynamics coupled to a discrete element method (CFD-DEM) was applied to simulate a pseudo-two-dimensional vibro-fluidized bed using \textit{cfdemCoupling}\textsuperscript{\tiny\textregistered} \citep{RN6}.

DEM was proposed initially by \cite{RN1803} and uses a Lagrangian description of the particle phase. The trajectories of the individual particles are determined by the action of interparticle contact forces $\mathbf{F}_{c,j}$, torques $\mathbf{M}_{c,j}$, gravity $\mathbf{g}$ and the drag force $\mathbf{F}_{d,j}$ exerted by the fluid flow around the particles \citep{RN1802,RN1801}. Momentum equations for the translational and rotational motion were solved for each particle $j$ of mass $m_j$ and diameter $d_j$ viz.:  
\begin{align}
    m_j\frac{\mathrm{d}^2\mathbf{x}_j}{\mathrm{d} t^2} &= \mathbf{F}_{c,j}+\mathbf{F}_{d,j} +m_j\mathbf{g}, \\
    \frac{1}{10}m_jd_j^2 \frac{\mathrm{d}\bm{\omega}_j}{\mathrm{d}t}&=\mathbf{M}_{c,j},
\end{align}Here, $\mathbf{x}_j$ denotes the position and $\bm{\omega}_j$ the angular velocity of the particle. Details on the implemented contact laws for $\mathbf{F}_{c,j}$ and $\mathbf{M}_{c,j}$ are described by \cite{RN6}. The gas flow (CFD, Eulerian description) was modelled by an incompressible, two-phase formulation of the Navier-Stokes equations including a term $\mathbf{R}$ to capture the momentum exchange between the gas phase and the suspended particles \citep{RN1801}: 
\begin{align}
    \frac{\partial \epsilon}{\partial t} +\nabla \cdot\left( \epsilon \mathbf{u}\right) &= 0, \\
    \frac{\partial \left(\epsilon\mathbf{u}\right)}{\partial t} +\nabla \cdot\left( \epsilon \mathbf{u}\mathbf{u}\right) &= -\epsilon \frac{\nabla p}{\rho_g} - \frac{\mathbf{R}}{\rho_g} + \epsilon \nu \nabla^2 \mathbf{u} + \epsilon \mathbf{g}, \label{eq:Momentum}
\end{align}where $\mathbf{u}$ is the gas velocity, $p$ is the gas pressure, $\rho_g$ is the gas density, $\nu$ is the viscosity and $\epsilon$ is the local void fraction. $\epsilon$ was determined using the positions of the particles. The momentum exchange $\mathbf{R}$ between the  fluid and the particles inside a finite volume element $\Delta V$ was calculated as
\begin{equation}
     \mathbf{R} = \frac{\Sigma_{j = 1}^n \mathbf{F}_{d,j}}{\Delta V}.   
\end{equation}Here, the drag force was described by the Koch-Hill model \citep{RN7, RN1808}. The gas flow field was calculated in cubic volume elements of 5~mm edge length using the built-in finite volume method solver of \textit{cfdemCoupling}\textsuperscript{\tiny\textregistered} \citep{RN6}. The boundary conditions used in the CFD are summarized in table~\ref{tab:table2}. Due to the coarse resolution of the CFD mesh, a full-slip boundary condition was implemented between solid walls and the gas phase \citep{RN1805, RN1806, RN1807}. However, the frictional and collisional contacts between the particles and the walls are fully resolved in the DEM.

\begin{table}
\begin{center}
\def~{\hphantom{0}}
\begin{tabular}{llccc}
 & & \multicolumn{3}{c}{\textbf{Boundary}}\\
\textbf{Quantity}&  &\textbf{Inlet}& \textbf{Outlet}& \textbf{Walls}\\[3pt]
Reduced pressure & $p/\rho_g$ & zero gradient & $10^5 \text{m}^2/\textrm{s}^2$ & zero gradient \\
Gas velocity & $\mathbf{u}$ & $(0, U, 0)^T$ & zero gradient & full slip
\end{tabular}
\caption{ Boundary conditions used in the CFD part.} \label{tab:table2}
\end{center}
\end{table}

The numerical setup of the vibro-fluidized bed resembled the experiments of \citet{RN5}, see figure~\ref{fig:TimeSeries}~(a). Specifically, a box of width $W = 200 \text{ mm}$ and thickness $T = 10 \text{ mm}$ was filled up to a height of $H$~=~250~mm with a polydisperse mixture of spherical particles, referred to as the bulk phase. The bulk particles had a density $\rho_b = 6000 \text{ kg/m}^3$ and a mean diameter $d_b = 1.16 \text{ mm}$. Polydispersity was introduced to reduce crystallization effects of the packing \citep{RN1804}. Inside the bulk phase, a square cuboid of width $W_c = 30\text{ mm}$ and depth $T$ was cut out and filled with spherical particles of mean diameter $d_c$ and density $\rho_c$ (referred to as granular cluster). The lower edge of the cluster was 30~mm above the bottom of the bed. Details on the particle properties are found in table~\ref{tab:table1}. At the bottom of the bed, an upwards gas flow with a superficial velocity $U = 1.13 \text{ m/s}$, density $\rho_g = 1.2 \text{ kg/m}^3$ and viscosity $\nu = 1.5\times10^{-5} \text{ m}^2/\text{s}$ entered the bed. The value of $U$ was equal to the minimum fluidization velocity of the bulk particles ($U_{mf,b}$), i.e. the velocity at which the drag force balances the weight of the bulk particles in the gravitational field $g = 9.81 \text{ m/s}^2$ \citep{RN8}. Fluidization reduced frictional forces between the bulk particles, a key requirement for granular clusters being able to rise/sink. During the simulations, the box was subjected to a vertical, sinusoidal vibration with an amplitude $A = 1 \text{ mm}$ and frequency $f = 10 \text{ Hz}$. Applying vibration reduced the emergence of gas bubbles that otherwise arise when the particles are fluidized by gas alone \citep{RN5,RN9}. This allowed an unimpaired motion of the granular clusters. Importantly, applying only vibration did not establish any convective flow patterns in the bulk phase.

\begin{table}
\begin{center}
\def~{\hphantom{0}}
\begin{tabular}{p{4cm}p{3cm}p{3cm}}

\textbf{Quantity}& \textbf{Bulk particles}& \textbf{Cluster particles}\\[3pt]
Particle shape & Sphere & Sphere \\
Mean particle size & $d_b = 1.16$ mm & $d_c = 0.5 ... 3\times d_b$\\
Particle size distribution & $\Delta Q_3(d_b) = 0.6$ & $\Delta Q_3(d_c) = 0.6$ \\
  (mass fraction) & $\Delta Q_3(0.9d_b) = 0.2$ & $\Delta Q_3(0.9d_c) = 0.2$ \\
  & $\Delta Q_3(1.1d_b) = 0.2$ & $\Delta Q_3(1.1d_c) = 0.2$ \\
  Particle density & $\rho_b = 6000 \text{ kg/m}^3$ & $\rho_c = 0.2 ... 2 \times \rho_b$ \\
  Young's modulus & $E_b= 5 $ MPa  & $E_c= 5$ MPa  \\
  Poisson ratio & $\nu_b= 0.45 $ & $\nu_c= 0.45 $ \\
  Coefficient of restitution & $e_b= 0.3 $ & $e_c= 0.3 $ \\
  Coefficient of friction & $\mu_b= 0.15 $ & $\mu_c= 0.15 $ \\
  Contact law & Hertzian & Hertzian \\
\end{tabular}
\caption{Particle properties used in the DEM part.} \label{tab:table1}%
\end{center}
\end{table}

\section{Derivation of the analytic gas shift model}
\label{sec:Gasshift}

\begin{figure}
\centerline{\includegraphics[width=\linewidth]{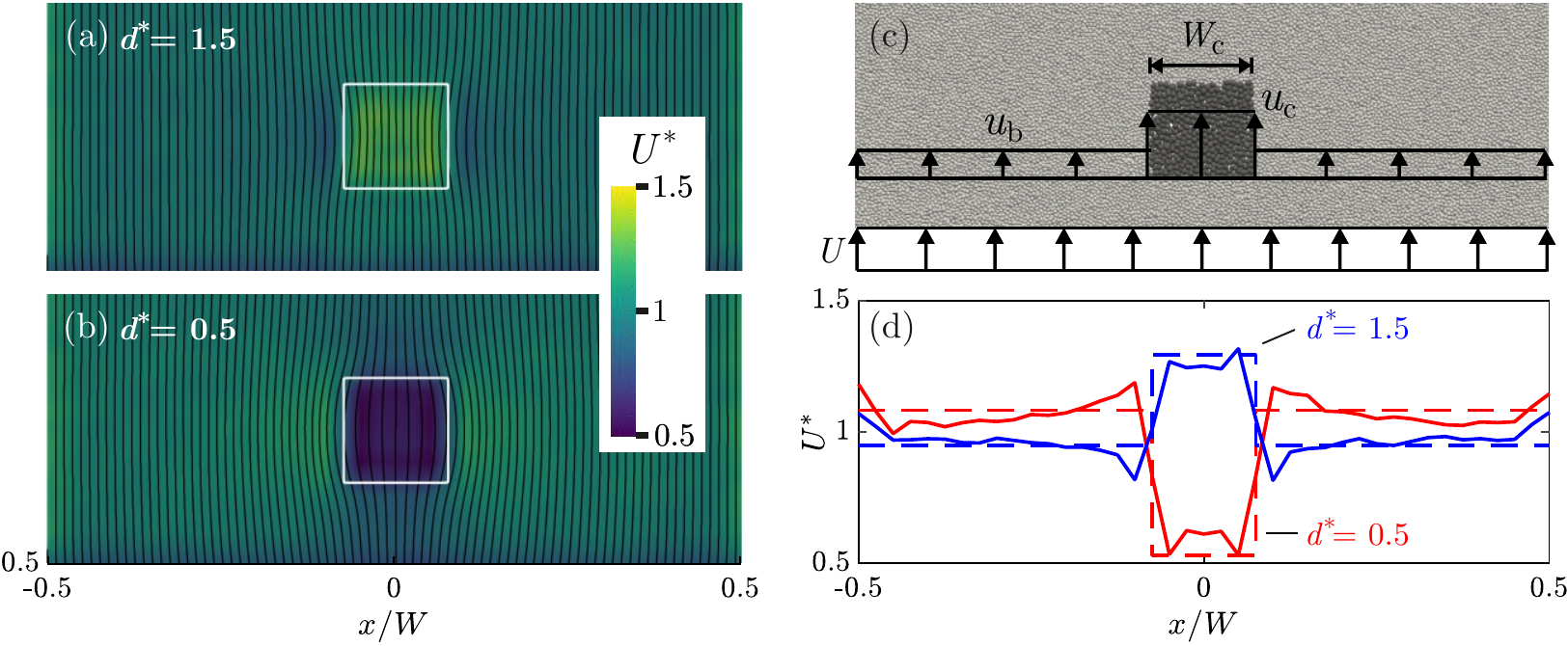}}
\captionsetup{width=\linewidth}
\caption{\label{fig:GasPhase} Heterogeneous gas flow near a square-shaped granular cluster of width $W_c = 30 \text{ mm}$. (a) and (b) Simulated gas flow with $\Rey_b = Ud_b/\nu = 88.75$ through a packing with $d^* = 1.5$ and 0.5, respectively. The white boxes mark the edge of the granular cluster and the black curves show the gas streamlines. $U^*$ is given by background colour. (c) Decomposition of $U$ into $u_c$ (cluster) and $u_b$ (bulk phase) according to Eqs.~(\ref{eq:Conti}) and (\ref{eq:Pressure}). (d) $U^*$ along the horizontal line through the centre of a granular cluster with, respectively, $d^* = 1.5$ (blue) and $d^* = 0.5$ (red). Solid lines plot the Eulerian-Lagrangian simulation results; dashed lines plot the solutions of the analytical model. }
\end{figure}

First, we evaluate numerically the effect of a granular cluster on the gas flow field. Figure~ \ref{fig:GasPhase} (a) and (b) display the gas flow in the vicinity and inside a granular cluster for two values of the relative particle diameter $d^* = d_c/d_b$. The black curves represent the gas streamlines and the background shows the dimensionless magnitude of the gas velocity $U^* = u/U$, where $u$ is the magnitude of the local superficial gas velocity. For $d^* = 1.5$, the streamlines concentrate inside the granular bubble and $U^*$ is higher in the cluster than in the surrounding bulk phase, i.e. gas flows preferentially through the granular cluster. For $d^* = 0.5$ the situation inverts. Such flow characteristics have been expected by \citet{RN10} and \citet{RN5} due to an increased/reduced permeability with increasing/reduced particle size \citep{RN11,RN12,RN13}. The flow heterogeneity that is caused by the redirection of gas into or around a granular cluster will be referred to as "gas shift" in the following. This gas shift is key to explain the rising and sinking of granular bubbles and droplets and hence has to be incorporated in any predictive analytical model.

In order to quantify the gas shift, we derive an analytical model for the gas flow around the granular cluster. For the model we apply a horizontal cut through the centre of the granular cluster that divides the bed into two regions: the cluster of dimensionless width $W^* = W_c/W$ and the bulk phase of dimensionless width $(1-W^*)$ as shown in figure~\ref{fig:GasPhase} (c). The gas that enters the bed is decomposed into two parallel flows through these two regions. Gas is assumed to flow with a uniform dimensionless velocity $U_c^* = u_c/U$ and $U_b^* = u_b/U$ through the cluster and bulk phase, respectively. Due to its low velocity, the gas flow is assumed to be incompressible and the continuity equation reads
\begin{equation} \label{eq:Conti}
    1 = (1-W^* ) U_b^*+W^* U_c^*.
\end{equation}
Figure~\ref{fig:GasPhase} (a) and (b) show parallel and equally spaced streamlines below and above the granular cluster, i.e. no horizontal pressure gradients exist in these regions. This implies that the gas flow through the cluster is driven by the same vertical pressure drop as the flow through the adjacent bulk phase. Ergun’s equation \citep{RN11} is used to quantify the pressure drop  $\Delta p$ across a height $L$ of cluster or bulk phase particles: 
\begin{equation}  \label{eq:Ergun}
    \frac{\Delta p_i}{L}=\frac{150\mu_g}{d_i^2}\frac{(1-\epsilon_i)^2}{\epsilon_i^3}u_i+\frac{1.75\rho_g}{d_i}\frac{1-\epsilon_i}{\epsilon_i^3} u_i^2,
\end{equation}where index $i = [b,c]$ denotes the bulk or cluster phase, respectively, and $\mu_g = \nu \rho_g$ is the dynamic gas viscosity. Equating $\Delta p_c/L = \Delta p_b /L$ and rewriting the equation in a non-dimensional form yields a correlation between $U_b^*$, $U_c^*$, $d^*$, the void fraction $\epsilon$ and the Reynold’s number of the bulk phase $\Rey_b = Ud_b/\nu$, see appendix \ref{app:A} for the full derivation. According to our simulations, $\epsilon$ is almost constant ($\epsilon \approx 0.4$) for a large range of $d_c$ and $d_b$, and we obtain
\begin{equation} \label{eq:Pressure}
    0=\left(U_b^*-\frac{U_c^*}{d^{*^2}}\right)+\frac{\Rey_b}{k}\left(U_b^{*^2}-\frac{U_c^{*^2}}{d^*} \right)
\end{equation}
with $k = 85.7(1-\epsilon) \approx 51.42$. Combining Eqs. (\ref{eq:Conti}) and (\ref{eq:Pressure}) allows us to determine $U_c^*$ and $U_b^*$ as a function of $\Rey_b$, $d^*$ and $W^*$. Figure~\ref{fig:GasPhase} (d) compares the values of $U^*$ along a horizontal line through the centre of a cluster obtained by the Eulerian-Lagrangian simulation (solid lines) with the analytical model (dashed lines). For both values of $d^*$, the analytical model gives an accurate quantitative prediction of the average gas velocity inside the granular cluster and the bulk phase, although not all local effects (e.g. wall effects at $|x/W| = 0.5$ or local depletion/accumulation of gas flow in the bulk phase for $0.075\leq |x/W|\leq 0.16$ due to the gas shift) are captured in the analytical model. Based on this finding, $U_c^*$ can be used as a quantitative measure for the gas shift as $U_c^*>1$ indicates an increased gas flow inside the granular cluster compared to the inlet gas velocity $U$. Inversely, $U_c^* <1$ denotes a reduced gas flow in the cluster. A further evidence for the validity of the analytical gas shift model is presented in figure~\ref{fig:GasShift}~(a) plotting $U_c^*$ vs.~$d^*$. Here, the predictions of the gas shift model agree very well with the results of the Eulerian-Langrangian simulations ($W^* = 0.15$). The analytical model further indicates that $U_c^*$ depends on the dimensionless width of the cluster $W^*$. An increase in $W^*$ yields higher values of $U_c^*$ for $d^* < 1$ and lower values for $d^* > 1$. This is because for a fixed $W$ an increasing width of the granular cluster reduces the size of the bulk phase from which gas is withdrawn ($d^* > 1$) or into which additional gas is shifted to ($d^* < 1$): thus, $U_c^* \rightarrow 1$ for $W^* \rightarrow 1$. Inversely, $U_c^*$ approaches a finite value for $W^*\rightarrow 0$. This solution represents the maximal obtainable gas shift to/from an infinitely small cluster for a given $d^*$ and $\Rey_b$.

\section{Derivation of the neutral buoyancy limit}
\label{sec:NeutralB}

\begin{figure}
\centerline{\includegraphics[width=\linewidth]{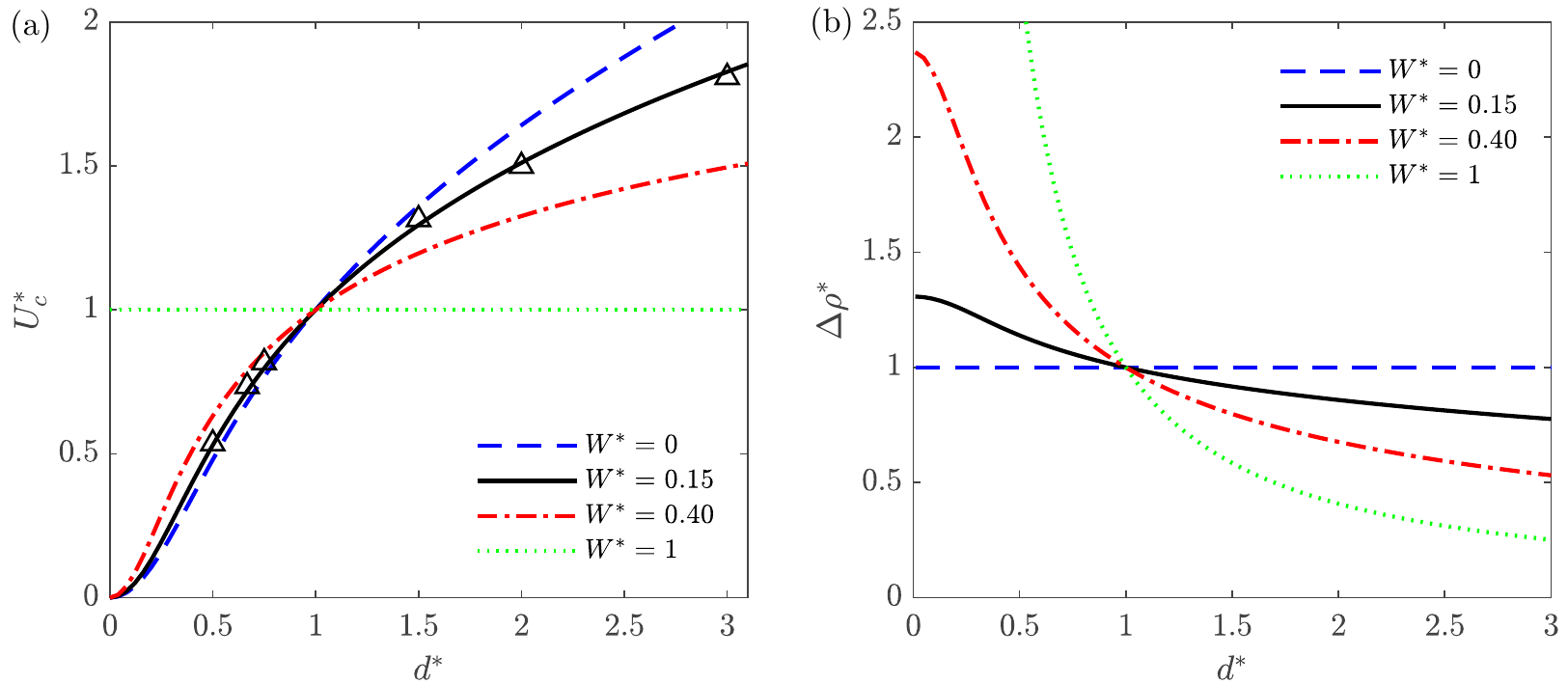}}
\captionsetup{width=\linewidth}
\caption{\label{fig:GasShift} (a) Gas shift ($U^*_c$) occurring in granular clusters of particle size $d^*$ as predicted by Eqs.~(\ref{eq:Conti}) and (\ref{eq:Pressure}) for a series of widths $W^*$. Triangles ($\vartriangle$) represent the results of the Eulerian-Lagrangian simulations for $U_c^*$ with $W^* = 0.15$. (b) Neutral buoyancy limits predicted for granular clusters of varying $W^*$ (Eqs.~(\ref{eq:Conti}-\ref{eq:Umf})) with $\Delta\rho^* = \left(\rho_c -\rho_g\right)\left(\rho_b - \rho_g \right)^{-1}$. Both panels (a) and (b) use $\Rey_b = 88.75$ and $\Arch_b = 3.40\times10^5$.}
\end{figure}

As the analytical model is able to accurately predict the gas shift to/from the granular cluster to the bulk phase, we address next the question under which circumstances a granular cluster rises or sinks. In our model we use two assumptions: first, the inlet gas velocity $U$ is set to the minimum fluidization velocity of the bulk particles $U_{mf,b }$, i.e. the velocity of the gas flow required to induce a drag force that equals the weight of the particles. $U_{mf,i}$ is determined from a force balance equating the pressure drop across the packing (Eq. (\ref{eq:Ergun})) with the weight of the packing per cross-sectional area such that 

\begin{equation}
 \label{eq:Umf_basis}
\frac{\Delta p_i}{L} =\frac{150\rho_g\nu}{d_i^2}\frac{(1-\epsilon)^2}{\epsilon^3}U_{mf,i}+\frac{1.75\rho_g}{d_i}\frac{1-\epsilon}{\epsilon^3} U_{mf,i}^2 = (1-\epsilon)(\rho_i -\rho_g)g,
\end{equation}where $i = [b, c]$\citep{RN8}. Rearranging the right hand sides and introducing the Reynolds number $\Rey_{mf,i}=d_i U_{mf,i} \nu^{-1}$ and the Archimedes number $\Arch_i = d_i^3(\rho_i-\rho_g)(\rho_g\nu^2)^{-1}$ yields

\begin{equation}\label{eq:Umf} 
\frac{1.75}{\epsilon^3}\Rey_{mf,i}^2+ \frac{150(1-\epsilon)}{\epsilon^3} \Rey_{mf,i} = \Arch_i.
\end{equation}Thus, for constant $\epsilon$ and $\nu$, $\Rey_{mf,b}$ and $U_{mf,b}$ are only functions of $\Arch_b$ and $d_b$. Applying this assumption to the bulk particles ($\rho_b = 6000 \text{ kg/m}^3$, $d_b = 1.16 \text{ mm}$, $\epsilon = 0.4$) and air ($\rho_g = 1.2 \text{ kg/m}^3$, $\nu = 1.5\times10^{-5} \text{ m}^2\text{/s}$) yields $U = U_{mf,b} = 1.13 \text { m/s}$.

Secondly, we assume that the rising/sinking of a cluster is driven by the degree of fluidization of the cluster particles due to the gas flow. This assumption is based on findings by \citet{RN14} and  \citet{RN15, RN16, RN17} who have studied size separation in binary granular materials that have been subjected to combined vibration and fluidizing gas flow. Their research revealed a competition between vibration-induced size separation as well-known from the Brazil nut effect \citep{RN1634,RN1774,RN1607} and its reverse \citep{RN1606,RN1604,RN1602} and gas-flow-induced size separation due to differences between the drag and gravitational force acting on the particles. They found that if the gas velocity $U$ is close to or above the minimum fluidization velocity of a granular material, size separation is governed by gas flow effects even for strong vibrational accelerations $\Gamma = A(2 \pi f)^2/g > 1$. Under such boundary conditions, binary granular materials segregate according to their $U_{mf}$ values with the material of lower $U_{mf}$ segregating on top of the material with the higher $U_{mf}$. In our studies, we use $\Gamma = 0.4$ and $U=U_{mf,b}$, i.e. size separation is dominated by gas flow effects. Hence, we hypothesize that a granular bubble rises, if the gas flow through the granular bubble gives rise to a drag force that is larger than the particle weight ($u_c > U_{mf,c}$) and the granular droplet sinks for $u_c < U_{mf,c}$; a granular cluster is neutrally buoyant if $u_c = U_{mf,c}$. From the analytical gas shift model we have $u_c = U U_c^*$ and $U_{mf,c}$ can be calculated via Eq. (\ref{eq:Umf}). Although vibration may reduce $U_{mf}$ of granular particles for $\Gamma \geq 1$ \citep{RN18,RN19} and $\Gamma < 1$ \citep{RN9}, we assumed that vibration does not affect appreciably the neutral buoyancy limit of the granular cluster because the net drag that leads to the rise of a granular bubble is dominated by the gas flow. Small fluctuations in the drag force due to vibration are expected to cancel out over a full vibration cycle. 

With these two assumptions at hand, the neutral buoyancy limit can be calculated as a function of $d^*$, $W^*$, $\Rey_b$ and $\Arch_b$. Figure~\ref{fig:GasShift}~(b) plots the relative density difference $\Delta\rho^* = \left(\rho_c -\rho_g\right)\left(\rho_b - \rho_g \right)^{-1}$ required to establish a neutrally buoyant granular cluster of width $W^*$ as a function of $d^*$ at constant $U$ (later referred to as "neutral buoyancy limit"). We observe that $\Delta\rho^*$ decreases monotonically with increasing $d^*$. For $d^* < 1$, the density of the cluster particles must be larger than the density of the bulk particles to achieve neutral buoyancy, and vice versa for $d^* > 1$. The monotonic behavior is due to the particle size dependence of $U_{mf,c}$. This dependence is depicted for $W^* = 1$, i.e. the entire bed is filled with cluster particles and no gas shift occurs ($U_c^* = 1$). Here, a constant $U$ leads to $\Delta\rho^* \rightarrow \infty$ (0) when $d^* \rightarrow 0$ ($\infty$). For $W^* < 1$, the $\Delta\rho^*$-curves increasingly flatten with decreasing $W^*$ due to the gas shift to/from the granular cluster. As confirmed in appendix \ref{app:B}, the density of the cluster particles must be equal to the density of the bulk particles to be neutrally buoyant for $W^*\rightarrow 0$ (dashed line in figure~\ref{fig:GasShift}~(b)). To summarize, our model predicts a granular bubble to rise if $\Delta\rho^*$ is below the neutral buoyancy limit for any given $\Arch_b$, $\Rey_b$, $W^*$ and $d^*$, otherwise a granular droplet sinks. 

\begin{figure}
\centerline{\includegraphics[width=\linewidth]{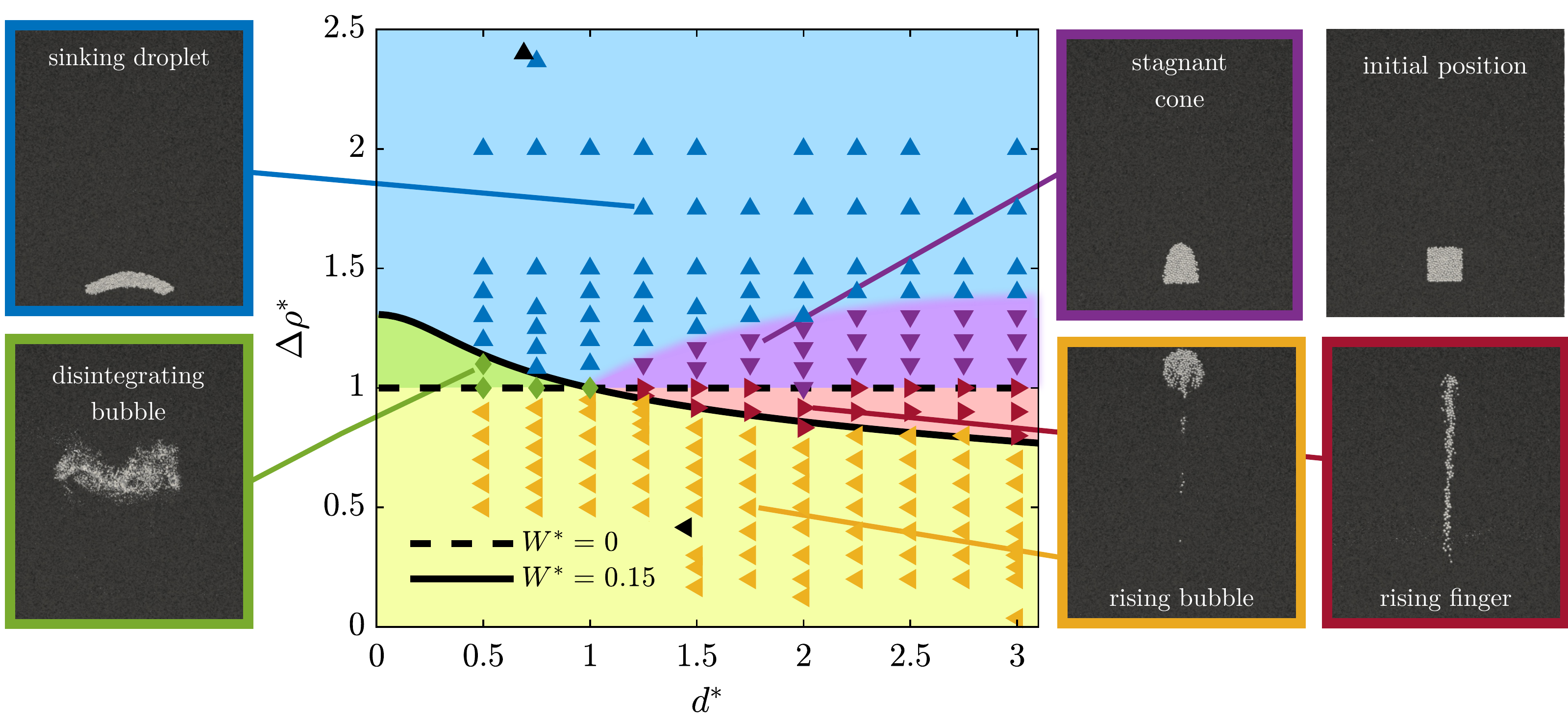}}
\captionsetup{width=\linewidth}
\caption{\label{fig:Phaseplot} Regime map for a granular cluster with $W^* = 0.15$, $\Rey_b = 88.75$ and $\Arch_b = 3.40\times10^5$. The solid and dashed lines are the neutral buoyancy limits for $W^* = 0.15$ and 0, respectively. The following regimes are observed: sinking droplet ($\vartriangle$, blue region), stagnant cone ($\triangledown$, purple region), rising finger ($\triangleright$, red region), rising bubble ($\triangleleft$, yellow region) and disintegrating bubble ($\Diamond$, green region). Coloured symbols are results of the Eulerian-Lagrangian simulations. The black filled symbols are experimental results from \citet{RN5}.}
\end{figure}

\begin{figure}
\centerline{\includegraphics[width=\linewidth]{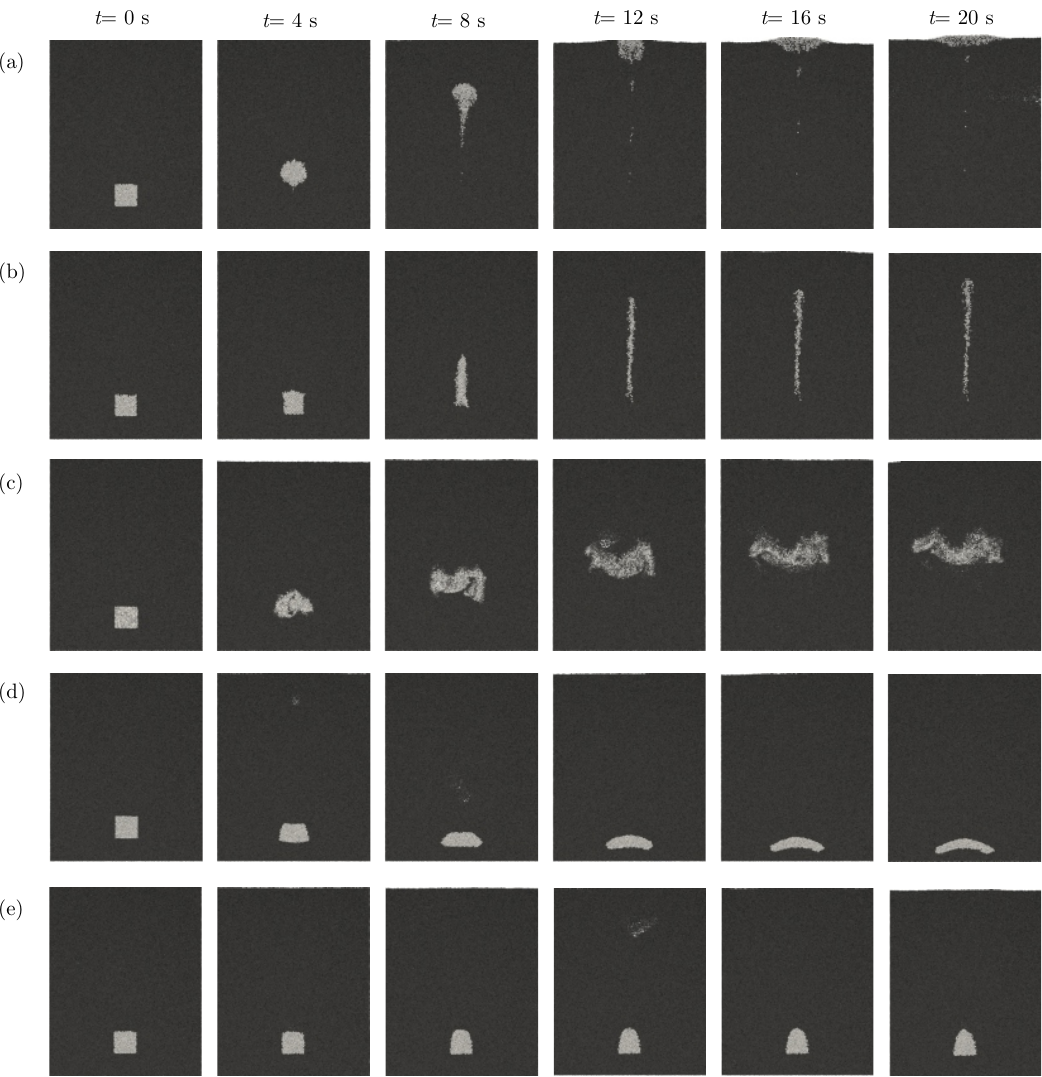}}
\captionsetup{width=\linewidth}
\caption{\label{fig:ClusterEvolution} Evolution of granular clusters (white) in a vibro-fluidized bed of bulk particles (black) for various values of $d^*$ and $\Delta\rho^*$. All simulations used $W^*=0.15$, $ \Rey_b = 88.75$ and $\Arch_b=3.40\times10^5$. (a) Rising bubble: $d^*=1.75$, $\Delta\rho^*=0.5$. (b) Rising finger: $d^*=2$, $\Delta\rho^*=0.9$. (c) Disintegrating bubble: $d^*=0.5$, $\Delta\rho^*=1.1$. (d) Sinking droplet: $d^*=1.25$, $\Delta\rho^*=1.75$. (e) Stagnant cone: $d^*=1.75$, $\Delta\rho^*=1.2$. }
\end{figure}

\section{Construction of a regime map}

The analytical model at hand, we constructed a regime map that predicts the rising or sinking of a granular bubble/droplet of a given size $W^*$ (figure~\ref{fig:Phaseplot}). In order to test the validity of the regime map, numerical simulations were performed to probe the $\Delta\rho^*$-$d^*$ space for a fixed cluster size of $W^* = 0.15$, $\Rey_b = 88.75$ and $\Arch_b = 3.40\times10^5$. Examples of such simulations are depicted in the time series of figure \ref{fig:ClusterEvolution}. The occurring motion patterns of the different granular clusters simulated were classified and marked in the regime map (figure \ref{fig:Phaseplot}). Five different regimes were identified: rising bubbles, rising fingers, disintegrating bubbles, sinking droplets, and stagnant cones. The previous experimental study of \citet{RN5} has identified only the rising bubble and sinking droplet regimes. As can be seen in the regime map, the first four regimes are in agreement with our neutral buoyancy model and their regime boundaries coincide with the neutral buoyancy limit for $W^* = 0.15$ and $W^* = 0$ (i.e. $\Delta\rho^* = 1$). 

Rising bubbles occur if $\Delta\rho^*$ is below the neutral buoyancy limit of the bubble for $d^* \geq 1$ or below $\Delta\rho^* = 1$ for $d^* < 1$ ($\triangleleft$, yellow region in figure~\ref{fig:Phaseplot}). In this region, the cluster particles form a roundish bubble that rises to the top of the bed. For clusters with $d^* < 1$ and $\Delta\rho^*$ close to the neutral buoyancy limit, the rising bubbles are elongated and leave behind a trail of particles in the bulk phase (figure \ref{fig:ClusterEvolution}~(a)). A decrease in $\Delta\rho^*$ reduces this particle trail and the bubbles become cap-shaped as can be seen in figure \ref{fig:TimeSeries}~(b). 

Rising fingers occur for clusters with $d^* > 1$ and $\Delta\rho^*$ between the neutral buoyancy limit and $\Delta\rho^* = 1$ ($\triangleright$, red region). In this regime, the cluster strongly elongates and only a thin stream of particles, termed rising finger, rises through the bulk phase. An example thereof is given in figure \ref{fig:ClusterEvolution}~(b). The finger is thinnest for $\Delta\rho^*$ close to unity and becomes thicker as $\Delta\rho^*$ approaches the neutral buoyancy limit ($W^* = 0.15$). Our analytical model explains the emergence of rising fingers: If $\Delta\rho^*$ is larger than the neutral buoyancy limit, $u_c$ is insufficient to levitate the entire granular cluster. However, small disturbances decrease the cluster width and thereby increase the gas flow inside the finger (see figure~\ref{fig:GasShift}~(a)). Thus, the rising finger is stabilized by decreasing its thickness.

In contrast to the self-stabilized rising fingers, disintegrating bubbles form for $d^* < 1$ and $\Delta\rho^*$ between unity and the neutral buoyancy limit ($\Diamond$, green region in figure~\ref{fig:Phaseplot}). Here, the cluster rises first as a coherent granular bubble, but disturbances during its rise induce its disintegration as seen in figure~\ref{fig:ClusterEvolution}~(c). According to our model, the granular bubble rises because $\Delta\rho^*$ is below the neutral buoyancy limit, but an increase in bubble width due to small disturbances increases the gas flow inside the bubble. The increased gas flow leads to the formation of small gas pockets within the granular bubble amplifying the effect of the disturbances and leading in turn to a full disintegration of the granular bubble.

For values of $\Delta\rho^*$ significantly larger than unity, a granular droplet sinks ($\vartriangle$, blue region in figure~\ref{fig:Phaseplot}), as the cluster particles are too heavy to be carried up by the drag exerted by the gas flow. During its descent, the sinking droplet flattens and performs ultimately a binary split (figure~\ref{fig:ClusterEvolution}~(d)). The splitting of such a granular droplet is not caused by the cluster's proximity to the bottom of the fluidized bed, but is a property that also exists for larger distances to the confining walls. \citet{RN5} discovered the splitting of these droplets, however the mechanism behind the droplet splitting has yet to be revealed. 

Although the analytical model predicts all droplets with $d^* > 1$ and $\Delta\rho^*>1$ to sink, there is a region for $\Delta\rho^*\geq1$ and $d^* > 1$, where droplets become stagnant and develop a conical shape, termed stagnant cone ($\triangledown$, purple region in figure~\ref{fig:Phaseplot}). The lower limit of the stagnant cone regime is at $\Delta\rho^* = 1$ and the upper limit is increasing with increasing $d^*$. As the stagnant cone particles are too heavy to be neutrally buoyant ($\Delta\rho^*>1$), the cluster acts as an additional load and inhibits fluidization of the bulk particles underneath the cluster. However due to the small values of $\Delta\rho^*$, the net gravitational force of the cluster cannot overcome the interparticle friction in this unfluidized bulk region and thus the cluster remains stagnant. The formation of a cone (figure~\ref{fig:ClusterEvolution}~(e)) can be explained by small fluctuations in the local gas velocity due to bed vibration. These fluctuations lead to a temporary fluidization of the cluster particles at the top of the cluster, such that they rearrange to an incipient granular finger. However, the fluctuations cannot maintain fluidization conditions for an extended duration of time, i.e. the cluster particles rest on top of the granular cluster. For $\Delta \rho^*\rightarrow1$, the cone sharpens further until it finally merges into a rising finger for $\Delta \rho^*=1$. 

We have performed additional numerical Eulerian-Lagrangian simulations in which we varied the cluster width and were able to confirm the validity of the neutral buoyancy limit for a series of $W^*$ = [0.1, 0.15, 0.3, 0.4]. However, the derivation of our model requires $d_c/W_c \ll 1$ to allow for enough particles in the granular cluster to justify the use of Ergun’s equation \citep{RN11}. Also, our model is most accurate for $0.4 < d^* < 2.5$, as outside these limits small particles percolate into the interstitial voids of the larger particles and change the permeability of the cluster packing.

\section{Conclusion}
\label{sec:conclusion}
We have investigated the physics controlling the motion of a granular cluster in a pseudo-two-dimensional vibro-fluidized bed. A granular cluster is an agglomeration of granular particles of certain size and density that is submersed in a bulk of particles with different particle size and/or density. 
When fluidized by combining vibration and gas flow, the granular cluster either rises coherently to the freeboard of the bed (granular bubble) or sinks to the bottom of the bed (granular droplet) depending on the relative size and relative density of the cluster particles and the bulk particles \citep{RN5}. This work investigated the transition between rising granular bubbles and sinking granular droplets and identified the conditions required to form neutrally buoyant granular clusters.
Eulerian-Lagrangian simulations revealed the existence of a gas flow heterogeneity in the proximity of a granular cluster, where gas flows preferentially through regions of larger particles due to their increased permeability. This shift in the gas flow pattern (into or out of the granular cluster) affects locally the degree of fluidization of the particles in and around the cluster and in turn significantly influences the motion of the granular cluster itself. Based on a dimensionless gas shift model, we proposed an analytical model predicting the neutral buoyancy limit of a granular cluster assuming that under this condition the local gas velocity matches the minimum fluidization velocity $U_{mf,c}$ of the cluster particles. If the gas velocity is higher or lower than $U_{mf,c}$, the cluster rises or sinks, respectively. Finally, a dimensionless regime map was compiled and tested against extensive Eulerian-Lagrangian simulations. This regime map correctly predicts the regimes of rising bubbles and sinking droplets and revealed three distinct, previously unreported, regimes in the transition region: rising fingers, disintegrating bubbles and stagnant cones.

\backsection[Funding]{This work was supported by the Swiss National Science Foundation (grant number 200020\_182692). }

\backsection[Declaration of interests]{The authors report no conflict of interest.}

\backsection[Data availability statement]{The data that support the findings of this study are available from the corresponding author upon request.}

\backsection[Author ORCID]{
J. P. Metzger, https://orcid.org/0000-0002-1326-8204;
L. Girardin, https://orcid.org/0000-0002-8017-2405;
N. A. Conzelmann,  https://orcid.org/0000-0002-5568-4863;
C. R. Müller,  https://orcid.org/0000-0003-2234-6902}

\appendix

\section{Derivation of the gas shift model} \label{app:A}
The continuity equation of an incompressible gasflow of constant flow rate $Q = U W T$ that splits into two parallel flows is given by
\begin{equation}
    Q/T = U W = (1-W_b ) u_b + W_c u_c.
\end{equation}
Division by $U W$ and introduction of $U_c^* = u_c/U$, $U_b^* = u_b/U$, and $W^* = W_c/W$ yields
\begin{equation}
    1 = (1-W^*) U_b^*+W^*  U_c^*. \label{apeq:conti}
\end{equation}
The pressure drop of a flow through a packing of height $L$ is given by \cite{RN11}
\begin{equation}
    \frac{\Delta p_i}{L}=\frac{150\mu_g}{d_i^2}\frac{(1-\epsilon_i)^2}{\epsilon_i^3}u_i+\frac{1.75\rho_g}{d_i}\frac{1-\epsilon_i}{\epsilon_i^3} u_i^2,
\end{equation}
where index $i = [b,c]$ denotes the bulk phase and the cluster phase (i.e. granular droplet/bubble), respectively, and $\mu_g = \nu \rho_g$ represents the dynamic gas viscosity. As $\Delta p_b/L = \Delta p_c/L$ and $\epsilon_b = \epsilon_c = \epsilon$, equating the pressure drop relations of the bulk and the cluster phase yields
\begin{equation}
    0= \left(u_b-\frac{u_c}{(d_c/d_b)^2}\right) +\frac{1.75d_b}{150(1-\epsilon)\nu} \left( u_b^2-\frac{u_c^2}{d_c/d_b }\right). 
\end{equation}
Now, we substitute $U_c^* = u_c/U$, $U_b^* = u_b/U$, $d^* = d_c/d_b$, $k = 150(1-\epsilon)/1.75$, and $\Rey_b = Ud_b/\nu$ to form dimensionless groups giving
\begin{equation}
    0=\left(U_b^*-\frac{U_c^*}{d^{*^2}} \right)+\frac{\Rey_b}{k} \left(U_b^{*^2}-\frac{U_c^{*^2}}{d^*} \right). \label{apeq:pressure}
\end{equation}
The combination of Eq.~(\ref{apeq:conti}) and (\ref{apeq:pressure}) determines the values of $U_c^*$ and $U_b^*$ for given $d^*$, $W^*$ and $\Rey_b$.

\section{Proof that $\Delta\rho^* = 1$ for $W^* \rightarrow 0$} \label{app:B}
Starting point of the neutral buoyancy limit for an infinitely small cluster ($W^* \rightarrow 0$) is the determination of the gas-shift in/around the granular bubble/droplet. The continuity equation reads $U_b^* = 1$ and the pressure relation yields
\begin{equation}
    0=\left( 1-\frac{U_c^*}{d^{*^2}} \right) +\frac{\Rey_b}{k} \left(1-\frac{U_c^{*^2}}{d^*} \right). 
\end{equation}
This expression is a quadratic equation that is solved for $U_c^*$ with the positive solution
\begin{equation}
    U_c^*=-\left(2k\Rey_b d^* \right)^{-1}+\sqrt{\left(2k\Rey_b d^* \right)^{-2}+d^*\left(k\Rey_b\right)^{-1}+d^* }. \label{apeq:Uc}
\end{equation}
Next, the minimum fluidization velocity of each particle type is determined by \cite{RN8}
\begin{equation}
    \frac{1.75}{\epsilon^3}\left( \frac{d_i U_{mf,i}}{\nu}\right)^2+ \frac{150(1-\epsilon)}{\epsilon^3} \left( \frac{d_i U_{mf,i}}{\nu} \right) = \frac{d_i^3(\rho_i -\rho_g)g}{\rho_g \nu^2},
\end{equation}
where we define $\Arch_i = \frac{d_i^3(\rho_i -\rho_g)g}{\rho_g \nu^2}$, $\Rey_i = \frac{d_i U_{mf,i}}{\nu}$ and $i = [b, c]$, i.e. 
\begin{equation}
    \frac{1.75}{\epsilon^3} \Rey_i^2+ \frac{150(1-\epsilon)}{\epsilon^3} \Rey_i  = \Arch_i.
\end{equation}
For $i = c$, this equation yields by substituting $\Rey_c = \Rey_b d^* U_c^*$

\begin{equation}
    \frac{1.75}{\epsilon^3} \Rey_b^2 \left(d^* U_c^* \right)^2 + \frac{150(1-\epsilon)}{\epsilon^3}  \Rey_b d^* U_c^*  = \Arch_c. \label{apeq:Re}
\end{equation}
Next, Eq.~(\ref{apeq:Uc}) and the definition of $k$ are substituted in Eq.~(\ref{apeq:Re}). Simplifying the expression results in
\begin{equation}
   d^{*^3} \left[\frac{1.75}{\epsilon^3}  \Rey_b^2+ \frac{150(1-\epsilon)}{\epsilon^3}  \Rey_b \right] = \Arch_c. \label{apeq:Ar}
\end{equation}
The bracketed term in Eq.~(\ref{apeq:Ar}) can be replaced by $\Arch_b$, yielding
\begin{equation}
    d^{*^3} \Arch_b= \Arch_c.
\end{equation}
Finally, re-substitution of the $\Arch_i$ definition and defining the relative density ratio to be 
\begin{equation}
    \Delta \rho^* = \frac{\rho_c -\rho_g}{\rho_b-\rho_g}
\end{equation}
leads to
\begin{equation}
    \rho_b-\rho_g = \rho_c - \rho_g \leftrightarrow \Delta \rho^* = 1.
\end{equation}

\bibliographystyle{jfm}
\bibliography{jfm}

\end{document}